\documentclass[preprint,epsfig,aps,superscriptaddress]{revtex4}
\usepackage{graphicx}
\begin{document}

\title{Orientational order in deposits of magnetic particles}

\author{J. M. Tavares}
\affiliation{Centro de F\'{\i}sica Te\'{o}rica e Computacional
da Universidade de Lisboa \\
Avenida Professor Gama Pinto 2, P-1649-003 Lisbon, Portugal}
\affiliation{Departamento de Ci\^{e}ncias Exactas e Tecnol\'{o}gicas,
Universidade Aberta \\
Rua Fern\~{a}o Lopes 9, $2^{\underline\circ}$ D$^{\underline{\rm to}}$,
P-1000-132 Lisbon, Portugal}
\author{M. Tasinkevych}
\affiliation{Departamento de Ci\^{e}ncias Exactas e Tecnol\'{o}gicas,
Universidade Aberta \\
Rua Fern\~{a}o Lopes 9, $2^{\underline\circ}$ D$^{\underline{\rm to}}$,
P-1000-132 Lisbon, Portugal}

\author{F. de los Santos}
\affiliation{Physics Department, Boston University \\
590 Commonwealth Avenue, Boston, MA 02215, USA}
\author{M. M. Telo da Gama}
\affiliation{Departamento de Ci\^{e}ncias Exactas e Tecnol\'{o}gicas,
Universidade Aberta \\
Rua Fern\~{a}o Lopes 9, $2^{\underline\circ}$ D$^{\underline{\rm to}}$,
P-1000-132 Lisbon, Portugal}
\affiliation{Departamento de F\'{\i}sica, Faculdade de Ci\^encias da
Universidade de Lisboa, \\
R. Ernesto Vasconcelos, Lisbon, Portugal}

\vfill
%\hfill{Typeset using REV\TeX}

\newpage

\begin{abstract}
We present preliminary results for the orientational order in
deposits of dipolar particles, on one dimensional substrates. 
The deposits are generated using a model where the incoming dipolar
particle interacts with the other particles in the deposit via a 
dipole-dipole potential. The interdipolar vectors are restricted to lie on
a square lattice although the dipole moments are free to rotate in three
dimensions.
The path of the incoming particle is generated through a Monte Carlo
scheme controlled by an effective temperature $T^*$, the case of pure
diffusion-limited deposition corresponding to $T^*\to\infty$. 

We calculate the ferromagnetic and nematic order parameters and the
dipolar orientational probability density of the deposits, at
various stages of growth and two effective temperatures. The dipolar
angular correlations along the rows and columns of the lattice are also 
investigated. 
We find that the orientational order of the deposits depends
strongly on the lattice structure, the stage of growth and the 
effective temperature.

\vspace{1cm}
\noindent PACS numbers: 82.20.Wt, 61.43.Hv, 64.60.Cn
\end{abstract}

\maketitle

\newpage

\section{Introduction}

\label{sec-intro}
The equilibrium properties of fluids formed by particles 
carrying strong permanent dipoles have been extensively studied in the 
last decade \cite{review}. 
The interest in these systems has grown due to the unusual results  
obtained in the numerical simulations of dipolar spheres. In fact, almost 
at the same time, Monte Carlo Canonical (MCC) simulations of dipolar hard 
spheres  in the canonical ensemble \cite{levesque:1993} and 
molecular dynamics simulation of dipolar soft spheres \cite{Patey:1992}, 
have shown that 
the dipolar fluids exhibit orientational 
(ferromagnetic) order at densities where the positional order 
is that typical of a liquid.      
Moreover, shortly after, MCC simulations  
revealed that the dipolar fluid  at very low densities 
does not behave as an ideal gas, 
but instead exhibits a complex structure, with the formation of chains 
and more complicated polidisperse 
aggregates \cite{levesque:1993b}. 
The correlations responsible by  these structures 
 have their  origin in the anisotropy and long 
range character of the dipolar potential that become 
relevant when the temperature is 
low enough (or the dipole moment is strong enough).
These striking results have led to theoretical developments
  that are capable to describe 
some of the peculiarities of dipolar fluids (see \cite{review} and 
references therein).   

The dramatic changes obtained for 
the equilibrium structure of simple fluids when dipolar interactions
are taken into account have motivated  us to investigate  the 
effects of this interaction in non-equilbrium processes. In  this work
we present preliminary results obtained when these interactions are 
included in a model of deposition  onto a 1-dimensional 
substrate.  

This paper is organised as follows: in section II we describe our model 
in detail, and relate it  with  some of 
the simplest  models of aggregation and deposition as well as with 
a recent study of 
aggregation of dipolar particles \cite{Rubi:1995}; in section 
III we present our preliminary results for the orientational properties 
of the dipoles in the deposits; finally, 
in section IV we summarise our findings.

\section{Model and simulation method}

\label{sec-meth}

One of the simplest models used to study cluster growth is DLA (Diffusion
Limited Aggregation) \cite{Meakinbook}. 
In this model particles are released, one by one, from a random position at 
a distance $R$ from a seed particle, and, after performing a random walk 
in $d$-dimensional space,  attach to the seed or to other particles that 
already part of the aggregate. DLA was generalized to  
study the growth of deposits on fibers or surfaces 
\cite{Meakinbook,Meakin:1983}. 
This generalization is the so called DLD (Diffusion Limited Deposition)
model. In DLD particles are deposited onto a $D$-dimensional substrate
($D=2$ for surfaces and $D=1$ for fibers). They diffuse through a $d=D+1$
dimensional space, starting at a random position a large distance 
from the substrate (a plane for $D=2$ and a line for $D=1$), and
eventually attach to the substrate or to the deposit formed by previously 
released particles.     
These simple models (and modifications thereof) were studied extensively 
during the last twenty years, with the goal of understanding the formation 
of complex patterns under conditions far from equilibrium  
({\it{e.g.}} \cite{Meakinbook,Barabasibook}). 
A particular extension of DLA was proposed recently in \cite{Rubi:1995}
where an off-lattice DLA model is generalized to include dipolar interactions 
between the particles. 
The dipolar interactions are taken into account through a Metropolis rule that 
changes the random diffusive movement of the particles. 
The interest in deposits of dipolar particles arises from the
strongly anisotropic character of the interactions, that may change the 
fractal dimension of the aggregates, at least under certain
conditions \cite{Meakinbook,Rubi:1995,jpcm:2002}. 
The model used in the present study \cite{jpcm:2002} 
can be viewed either as a generalization of DLD to include dipolar 
interactions or as an extension of the dipolar DLA to the growth of deposits 
on fibers.

The simulations were performed on a (1+1)-dimensional square lattice
of width $L=800a$ sites and any height to accommodate $M$ dipoles, 
where $a$ is the mesh spacing and the adsorbing substrate coincides with 
the bottom row (henceforth we take $a=1$). 
Periodic boundary conditions are imposed in the
direction parallel to the substrate. Each particle carries
a 3d dipole moment of strength $\mu$ and interacts through the pair
potential
\begin{equation}
\label{dipol_inter}
\phi_D(1,2) = -\frac{\mu^2}{r^3_{12}}
\left[ 3({\boldmath\hat\mu_1} \cdot {\boldmath\hat r_{12}}) 
({\boldmath\hat\mu_2} \cdot {\boldmath\hat r_{12}}) 
-{\boldmath\hat\mu_1} \cdot {\boldmath\hat\mu_2} \right], 
\qquad r_{12}\geq a,
\end{equation}
where $r_{12}$ is the distance between particles 1 and 2, 
${\boldmath\hat r_{12}}$ is the two-dimensional (2d) unit vector along 
the interparticle axis, and ${\boldmath\hat\mu_1}$ and 
${\boldmath\hat\mu_2}$ are the 3d unit vectors in the 
direction of the dipole moments of particles 1 and 2 respectively. 
Finally, `1' and `2' denote the full set of positional and 
orientational coordinates of particles 1 and 2.

A particle is introduced at a lattice site $(x_{in},H_{max}+A\,L)$,
where $x_{in}$ is a random integer in the interval $[1,L]$, $H_{max}$
is the maximum distance from the substrate to any particle in the 
deposit, and $A$ is a constant. The dipolar moment of the released 
particle is oriented at random. The particle then undergoes a random walk 
by a series of jumps to nearest-neighbour lattice sites, while
experiencing the dipole interactions with the particles
that are already attached to the deposit. We incorporate the 
effects of these interactions through the same  
Metropolis algorithm used in \cite{Rubi:1995}. If the 
deposit contains $M$ particles, then the interaction
energy of the $(M+1)$th incoming particle (the random walker) with the 
particles in the deposit is given by $E({\bf r},{\hat \mu})=\sum_{i=1}^M
\phi_D(i,M+1)$, where ${\bf r}$ and ${\hat \mu}$ are the current 
position and the dipole orientation of the random walker respectively
($ {\bf r} $ is a 2d vector). Then we randomly choose a new position 
${\bf r^{\prime}}$ ($|{\bf r} -{\bf r^{\prime}}| =a$) and a new dipole 
orientation ${\hat\mu}^{\prime}$ for the random walker \cite{footnote1}; 
this displacement 
is accepted with probability
\begin{equation}
\label{metropolis}
p={\rm min}\left\{1,\exp\left[- \frac{E({\bf r^{\prime}},{\hat
\mu^{\prime}})-E({\bf r},\hat \mu)}{T^*} \right]\right\}.
\end{equation}
$T^*=k_BTa^3/\mu^2$ is an effective temperature, 
inversely proportional to the dipolar energy scale. 
The long range of the dipole-dipole
interaction is taken into account in the calculation of 
the dipolar energy, through the 
evaluation of the Ewald sums for the particular geometry  
of this study: 3d dipoles in 2d space with periodic boundary conditions 
in one direction only \cite{jpcm:2002}.
In the limit 
$T^*\to \infty$ all displacements are accepted
and our model reduces to DLD. On the other hand, in the limit 
$T^*\to 0$ only displacements that lower the energy $E({\bf 
r},{\hat \mu})$ are accepted. 
Note that, in this limit, the model still has a certain degree of 
randomness as the $(M+1)th$ incoming particle may attach at a site which
is not necessarily that corresponding to the absolute minimum of the
energy (a lattice row or column).

The diffusion of a particle ends when it contacts
the deposit ({\it i.e.}, becomes a nearest neighbour of another particle that
is already part of the deposit) or the substrate ({\it{i.e.}}, 
its position has coordinates $(x,1)$).  
After contact, the particle's dipole reorients along the local field 
due to the other particles in the deposit.
When the distance of a particle from the substrate
exceeds $H_{max}+2A\,L$, the particle is removed from the system and
another particle is released. 
In the simulations reported here we took $A=1$; larger values of $A$ 
were tested and found to give the same results, but with drastically 
increased computation times.
  
Figure 1 exhibits snapshots of 
typical deposits obtained at high
temperature ($T^*=10^{-1}$ in black), and low temperature 
($T^*=10^{-4}$ in grey). 
Both deposits have the same general appearance, also observed in DLD:
they consist of several trees competing to grow. As the height of the
deposit increases, fewer and fewer trees `survive' ({\it{i.e.}}, carry on
growing), as a consequence of the so-called shielding or screening
effect. From figure 
1 this seems more pronounced at lower temperatures, since above 
a height of 1000 (about 1/8 of the maximum height of this deposit) 
only one tree survives. The study of the geometrical properties 
of these deposits was carried out elswhere \cite{jpcm:2002}. In this paper
we will concentrate on the study of the orientational 
properties of the dipolar deposits.

\section{Results and discussion}

\label{sec-res}

The orientational properties of the  
dipoles are studied for deposits at 
two different effective temperatures,
$T^*=10^{-1}$ and $T^*=10^{-4}$, for which we simulated 28
and 52 deposits respectively. These values of $T^*$ were chosen
since it is expected that their geometrical properties are close to those
of two limiting cases: DLD at the highest temperature, $T^*=10^{-1}$, and
$T^*=0$ at the lowest, $T^*=10^{-4}$ \cite{Rubi:1995,jpcm:2002}.

The study of the global orientational properties
of the dipoles as a function of the stage of growth, $M$, is
carried out through the calculation of the ordering
matrix ${\bf Q}$ and the first and second-rank order parameters,
$\langle P_1 \rangle$ and $\langle P_2 \rangle$, respectively (see
{\it{e.g.}} \cite{Patey:1992}). 
The  mean value of the elements of
${\bf Q}$ when the deposit has $M$ particles is
\begin{equation}
\label{eq:qnm} Q_{\lambda\nu}(M) 
= \langle\frac{1}{2M}\sum_{i=1}^M
 \left(3 \mu_\lambda(i) \mu_\nu(i)
-\delta_{\lambda\nu}\right)\rangle,
\end{equation}
where $\lambda$ and $\nu$ refer to the cartesian coordinates of the 
unit vector in the direction of the dipole moment and $\langle \ldots
\rangle$ stands for an average over all the deposits simulated. The $x$ 
and $y$ coordinates of the dipoles 
are parallel to the corresponding axes of the underlying square 
lattice.
In figure 2 we plot the values of $Q_{xx}$, $Q_{yy}$, $Q_{xy}$ and
$Q_{zz}$.   

$Q_{zz}$ approaches $-\frac{1}{2}$ very rapidly, corresponding to
the vanishing of the out of plane component of the
dipole moment, $\mu_z$: except for the first hundreds of particles, the
dipoles lie on the plane of the deposit. This property resembles that
of 2d dipolar fluids \cite{Weis:2002} and is due to the anisotropy of the
dipolar potential, that enforces the dipoles to lie on the plane of the
interdipolar vectors.
 
There is a very fast transient ($M < 30 - 100$ depending on
$T^*$) corresponding to deposits with a finite out of plane dipole
moment. In this initial regime $Q_{zz}$ decreases rapidly while $Q_{xx}$
increases and $Q_{yy}$ decreases. Although the
number of deposited particles is rather small the absolute value
of the $xx$ component of $\bf Q$ is systematically larger than that of the
$yy$ component suggesting an anisotropic initial growth.

We then distinguish an early stage ($30 < M < 100$ at $T^*=10^{-4}$ and
$100 < M < 3000$ at $T^*=10^{-1}$) where $Q_{xx}$ decreases while
$Q_{yy}$ increases steeply. This is followed by a late stage
characterized by slow increase of $Q_{xx}$ and a decrease of $Q_{yy}$ to
their asymptotic values of 0.25 (note that the horizontal scale of figure
2 is logarithmic).
$Q_{xx}$ exhibits a minimum at a value of M that depends on temperature
and $Q_{yy}$ has the opposite behavior with a maximum at (roughly) the
same value of M. The difference between these extrema decreases as $T$
increases. 

The value of $Q_{xy}$ is $\approx 0$ for the the entire deposit (even in
the initial stage) at both effective temperatures. In fact, all the the other 
non-diagonal elements, $Q_{\lambda \nu}$ with $\lambda \ne \nu$, also vanish 
and the ordering matrix is diagonal (coupling of the lattice and dipolar
potential anisotropies).

These results for $\bf Q$ imply that the orientation of the dipoles, 
after the first hundreds of particles are released, is restricted to the
plane of the deposit and is preferentially along one of the two
lattice directions: parallel to the substrate, at the initial stage of
growth, and perpendicular to it at the early stage. These
preferred directions result from the coupling between the lattice and
potential anisotropies: nearest neighbours on the lattice have horizontal
or vertical interdipolar vectors and the lowest energy of a dipole pair
corresponds to dipoles aligned along the interdipolar vector.  
Finally at the late stage of growth, the two lattice directions are
(almost) equivalent.   

The order parameters $\langle P_1 \rangle$ and $\langle P_2
\rangle$ provide information about the ferromagnetic and
nematic order of the system. 
In the simulations of liquid crystalline systems \cite{allenbook}
and dipolar fluids \cite{Patey:1992}, $\langle P_2 \rangle$ is identified 
with the the largest eigenvalue of the ordering matrix, $\Lambda_{max}$.
This definition is appropriate when the other eigenvalues are 
$\approx -\frac{1}{2}\Lambda_{max}$ \cite{allenbook}. 
As is clear from the results for $Q_{\lambda \nu}$, this condition is not
fullfilled in the present case. For instance, in the late stage of
growth, $Q_{xx}\approx Q_{yy} \approx \frac{1}{4}$, 
$Q_{zz}\approx=-\frac{1}{2}$ and $Q_{\lambda\nu}\approx 0$ if $\lambda
\ne \nu$. The eigenvalues $(\Lambda_{max}, \Lambda_{0}, \Lambda_{min})$
are then $(\frac{1}{4},\frac{1}{4}, -\frac{1}{2})$, 
meaning that the dipoles have a vanishing component along one direction
(perpendicular to the plane of the deposit) and that their orientations are 
distributed symmetrically along the other two perpendicular directions 
(within the plane of the deposit). Thus, the dipoles are ``randomly''
oriented in the plane of the deposit and the system has no orientational
order, apart from that imposed by the underlying lattice. 
This suggests an alternative definition of $\langle P_2 \rangle$ as
the difference between $\Lambda_{max}(M)$ and the maximum eigenvalue of 
$\bf Q$ obtained for a deposit of $M$ dipoles, randomly oriented on a
plane, $\Lambda_{max}^{ran}(M)$. 
The mean value of $\langle P_2 \rangle$ at each stage of the growth 
$M$ is then calculated from,
\begin{equation}
\label{eq:P2}
\langle P_2(M) \rangle= 
\langle \Lambda_{max}(M) \rangle - 
\langle \Lambda_{max}^{ran} (M)\rangle,
\end{equation}

where $\langle \Lambda_{max}^{ran} (M)\rangle$ was calculated using 
100 configurations of 50000 random dipoles.

Figure 3a shows the results for $\langle P_2\rangle$ at two temperatures. 
The nematic order parameter has the same qualitative behavior at 
the two temperatures. At $T^*=10^{-1}$ during the initial stage of growth  
$\langle P_2\rangle$ increases abruptly from zero to about 0.5 ($M <
100$). In the late stage ($M > 3000$) $\langle P_2\rangle$ oscillates 
around a small value ($<$ 0.1) and then decreases slowly to zero as $M
\to \infty$. The early stage ($100< M < 3000$) is dominated by a marked
decrease of $\langle P_2\rangle$ that goes almost to zero and then increases 
to a value below 0.1. 
The behaviour of $\langle P_2\rangle$ at the lower temperature is very
similar. 
 
Taking into account the results obtained for the elements of the ordering 
matrix ${\bf Q}$ this means that at the initial stage the dipoles have a 
tendency to point along the lattice direction parallel to the
substrate; at the early stage there is a tendency to align vertically; and 
finally, at the late stage of growth, there is a tendency to align both
vertically and horizontally with (almost) equal probability.
There are, however, quantitative differences between the behavior at the two 
temperatures.  
The tendency to align horizontally at the initial stage decreases with 
decreasing temperature, since as $T$ decreases it occurs for smaller
deposits and is characterized by a lower value of $\langle P_2\rangle$. 
On the other hand, the tendency to align vertically at the early stage 
is enhanced at low temperatures since as $T$ decreases the value of 
of the second maximum of $\langle P_2\rangle$ increases and
the decay towards the late stage behavior is much slower.

The ferromagnetic order parameter $\langle P_1 \rangle$ is defined, for 
equilibrium systems of dipolar particles, as the projection of the 
polarization onto the director of the system \cite{Patey:1992}. The
director, ${\hat d}$, is the eigenvector of $Q$ corresponding to 
$\Lambda_{max}$. Therefore, we calculate the ferromagnetic order parameter 
using,
\begin{equation}
\label{eq:P1}
\langle P_1(M) \rangle= 
\frac{1}{M}\langle \left| \sum_{i=1}^M {\hat \mu}_i \cdot 
{\hat d} (M) \right|\rangle. 
\end{equation}

Figure 3b shows the results for $\langle P_1(M) \rangle$ at two temperatures. 
In both cases $\langle P_1(M) \rangle$ reaches a maximum at the end of the
initial stage of growth and then falls rapidly in the early stage. At the
higher temperature $\langle P_1 \rangle$ is almost zero during the late
stage of growth. At the lower temperature, however, $\langle P_1 \rangle$ 
is non-zero in the late stage of growth with a broad maximum ($\sim
0.15$) at $M \sim 10000$. For larger $M$ $\langle P_1 \rangle$
exhibits a slight tendency to decrease but from these results it is not 
clear that the asymptotic value of the ferromagnetic order parameter
vanishes as $M \to \infty$. 
If we consider the results obtained previously for the elements of
the ordering matrix, we conclude that the increase in
ferromagnetic order for $M > 300$ (extending from the initial to
the late stage regimes) corresponds to dipoles aligned along the vertical
direction.

We proceed this preliminary analysis by calculating the 
orientational probability density (OPD) as a function of the stage of
growth, $P(\theta,$M$)$. $P(\theta,M)d\theta$ is the probability of finding
two dipoles with a relative orientation between $\theta$ and  $\theta
+d\theta$, in a deposit with $M$ dipoles. This is calculated by
dividing $[0;\pi]$ into intervals $\Delta \theta$, and counting the number
of pairs of dipoles $n(\theta_k,M)$, with relative orientation in the
range $]\theta_k;\theta_k +\Delta \theta[$, for a deposit with $M$
dipoles. The OPD is given by
\begin{equation}
\label{eq:opd}
P(\theta_k,M) = \frac{2}{M\left(M-1\right)}
\frac{\langle n(\theta_k,M)\rangle} {\Delta \theta}
\end{equation}
In figure 4 we plot the OPD at three stages of
growth, $M=2000$, 10000 and 50000, for $\Delta \theta
=\pi/100$. If the relative orientation of the dipoles were random
the OPD would be uniform ($1/\pi$). Our results indicate a highly
non-uniform behaviour. The peaks, at $0$, $\pi$, $\pi/2$, $\pi/4$ and $3\pi/4$,
occur at both temperatures, at every stage of growth. At the lower
temperature, the structure is enhanced and other peaks appear. The two
large peaks at $0$ and $\pi$ indicate that most
dipoles are aligned while the peak at $\pi/2$ shows that the
alignement occurs along two perpendicular directions.
At first sight, the OPD appears to be symmetric about
$\pi/2$. A more careful analysis, however, indicates a difference
in the heights of the peaks at $0$ and $\pi$, that is largest in the
early stage of growth and at the lowest temperature: the peak at $\pi$ is
lower than that at 0.
This indicates that the system exhibits ferromagnetic order. Consider an
off-lattice system where the OPD may be
approximated as $\langle P(\theta,M) \rangle \approx \frac{1}{\pi}
\left(1+A_1 P_1(\cos\theta) +A_2 P_2(\cos\theta)\right)$, 
with the coefficients $A_i$ proportional to 
the square of the order parameters $\langle P_i(M) \rangle$. 
This function has a maximum at $0$. If $A_1=0$ and 
$A_2\ne 0$ then it is symmetric about $\pi/2$; if, $A_1\ne 0$ 
the function has either a minimum or a (local) maximum 
at $\pi$, depending on the values of $A_i$; but, in any case, it 
is not symmetric about $\pi/2$. The difference between 
the values of the OPD at $0$ and $\pi$ is proportional to 
$\langle P_1\rangle^2$.     
We see from figure 4 that, although in the present case the OPD exhibits a
lot more structure due to the underlying lattice, the observed asymmetry
of the peaks at 0 and $\pi$ is consistent with the results for
$\langle P_1 \rangle$: at $T^*=10^{-1}$ the asymmetry is apparent at the 
early stage of growth, {\it{i.e.}}, for $M=2000$ only, while at
$T^*=10^{-4}$ the asymmetry is evident for all stages of growth.

The results for the OPD and the order parameter $\langle P_1\rangle$ 
may be compared with the results of \cite{Rubi:1995}. In this off-lattice 
study of  DLA-dipolar clusters, the OPD was calculated and the 
function $\frac{1}{\pi}(1+A_1(T^*) \cos\theta)$ was shown to be a 
good approximation to it. At the temperatures considered in this 
work the values $A_1(10^{-1})\approx 0$ and $A_1(10^{-4})\approx0.05$ 
were obtained in \cite{Rubi:1995}. 
Thus, our results are consistent with those of \cite{Rubi:1995}:
both dipolar DLA clusters and dipolar DLD deposits 
become polarized below a certain temperature, at least for values of $M
\sim 50000$.

To illustrate the relation between the results obtained so far and the
spatial properties of the deposits and to motivate our
last calculation, we plot in figure 5 two snapshots at
$T^*=10^{-4}$ and $T^*=10^{-1}$ (dipoles with $|\hat \mu_x|/|\hat \mu_y|<1$ 
($>1$) are plotted in black (grey)). 
These structures are formed mostly by horizontal grey rows and vertical
black columns, suggesting that the particles have a tendency to form 
chains, along the lattice directions.
Elsewhere, the connection between the spatial and the orientational order
was verified quantitatively \cite{jpcm:2002}. Briefly, it was shown that 
the orientation of a dipole on the substrate has a tendency 
(that increases with decreasing temperature) to be in the direction 
of growth of the deposit at the site where the dipole was attached.   
Here, we want to quantify the relation between the orientation of
a pair of dipoles and their distance along a chain, at different 
temperatures $T^*$.
Since the dipoles are oriented along the rows and columns of the
underlying lattice, in a head to tail alignment, an estimate of the
average size of the dipolar chains is given by the distance beyond
which the orientational correlations vanish. 
For a given deposit we calculate the average value of
the cosine of the angle between all pairs of dipoles along rows (or
columns) that are a given distance apart. For columns this is
\begin{equation}
\label{eq:cosalphv} 
\langle \cos \alpha(r) \rangle_{\rm v}=
\langle \sum_{i=1}^L \sum_{j=1}^{h^*(i)} \frac{\hat \mu(i,j)\cdot
\hat \mu(i,j+r)}{n_d(i,r)} \rangle,
\end{equation}
where $\hat \mu(i,j)$ is the dipolar unit vector at site $(i,j)$, 
$h^*(i)$ is the maximum height of column $i$ and
$n_d(i,r)$ is the number of pairs of dipoles in column $i$ a distance
$r$ apart. An analogous calculation for rows is,
\begin{equation}
\label{eq:cosalphh} 
\langle \cos \alpha(r) \rangle_{\rm h}=
\langle \sum_{j=1}^{h^*} \sum_{i=1}^{L} \frac{\hat \mu(i,j)\cdot
\hat \mu(i+r,j)}{n_d(j,r)} \rangle.
\end{equation}
where $h^*$ is the maximum height of the deposit, $n_d(j,r)$ is the
number of pairs of dipoles in row $j$ a distance $r$ apart. Periodic
boundary conditions in the lateral direction are taken into account.

Figure 6 shows $\langle \cos \alpha(r)
\rangle_{\rm v}$ and $\langle \cos \alpha(r) \rangle_{\rm h}$ at
the late stage of growth ($M=50000$). At short distances,  
the curves exhibit an oscillatory behaviour, along rows and columns, at
both temperatures. At intermediate distances the curves decrease
exponentially and at large distances they oscillate around zero 
(for the distances depicted in figure 6, this is seen only for the 
horizontal case at $T^*=10^{-1}$), corresponding to the loss of
correlation between the dipolar orientations.

The  oscillatory behavior at short distances is due to the 
interplay beween the anisotropy of the dipolar interaction and 
the lattice. 
Suppose that a dipole makes an angle $\alpha$ with {\it {e.g.}} the 
vertical direction. If another dipole attaches to the deposit as a 
top neighbour, the angle with the vertical that minimizes
the interaction energy between the two dipoles is $-\alpha$ (neglecting 
the interactions with the other dipoles in the deposit). It then follows
that the third, fifth, {\it {etc.}}, dipoles align at an angle $\alpha=0$ 
with the first giving rise to the (damped) oscillations observed at short
distances ($r < 5-20$ lattice spacings). 

The decay of angular correlations, for rows and columns, is temperature
dependent. At fixed $r$, 
$\langle \cos \alpha(r) \rangle_{\rm v}$ and $\langle \cos
\alpha(r) \rangle_{\rm h}$, are larger at the lower temperature, since
the tendency for dipolar alignment increases as the temperature is
lowered. In the inset of figure 6 we have plotted $\langle \cos \alpha(r)
\rangle_{\rm v}$ and $\langle \cos \alpha(r) \rangle_{\rm h}$ in a 
logarithmic scale. The plots are linear indicating that the correlations
decay exponentially. The differences are not large, but the decay of
correlations is faster at higher temperatures (for rows and columns) and
slower for columns (at any fixed temperature).
We conclude that the dipoles are aligned in columns or rows forming chains
with a typical size, that depends on temperature. A similar analysis at
the early stage of growth will probably reveal larger differences. 

\section{Concluding remarks}

\label{sec-disc}

From the results for the order parameters and the ordering matrix $\bf Q$
we conclude that the dipoles of the deposits have a slight tendency to
align parallel to the substrate at the initial stage and perpendicular to
it during the early stage of growth. The vertical (horizontal) alignment
is enhanced by decreasing (increasing) temperatures. We have found
evidence that the dipoles exhibit ferromagnetic order, at low
temperatures, in agreement with the results of the off-lattice dipolar DLA 
model \cite{Rubi:1995}. 
However, we cannot discard the possibility that this ferromagnetic order 
is a finite size effect: in fact, at $T^*=10^{-4}$ 
the mean height of the deposit for $M=50000$ is almost one order of 
magnitude larger than that at $T^*=10^{-1}$ \cite{jpcm:2002}.   

The fact that the nematic
director is always either horizontal or vertical is of course a
lattice effect; but the existence of a temperature dependent
nematic order that changes from parallel to the substrate (in the
initial) to perpendicular (in the early stage of growth), seems more
likely to occur in off-lattice versions of this model. 

Finally, we have shown that the chained structure of the 
deposits has a characteristic length that depends on the direction of the 
chains and on the temperature.
The relation between the orientational order studied in this work and the  
geometrical properties (fractal dimension) of the 
deposits \cite{jpcm:2002} is left for future work.

\section*{Acknowledgements}

Funding from the Funda\c{c}\~{a}o para  a Ci\^{e}ncia e
Tecnologia (Portugal) is gratefully acknowledged in the form of
post-doctoral fellowships nos.\ SFRH/BPD/5654/2001 (F. de los
Santos) and SFRH/BPD/1599/2000 (M. Tasinkevych) and a plurianual running
grant.

\newpage

\begin{figure}
\vspace*{2.5cm}
\includegraphics[width=7cm]{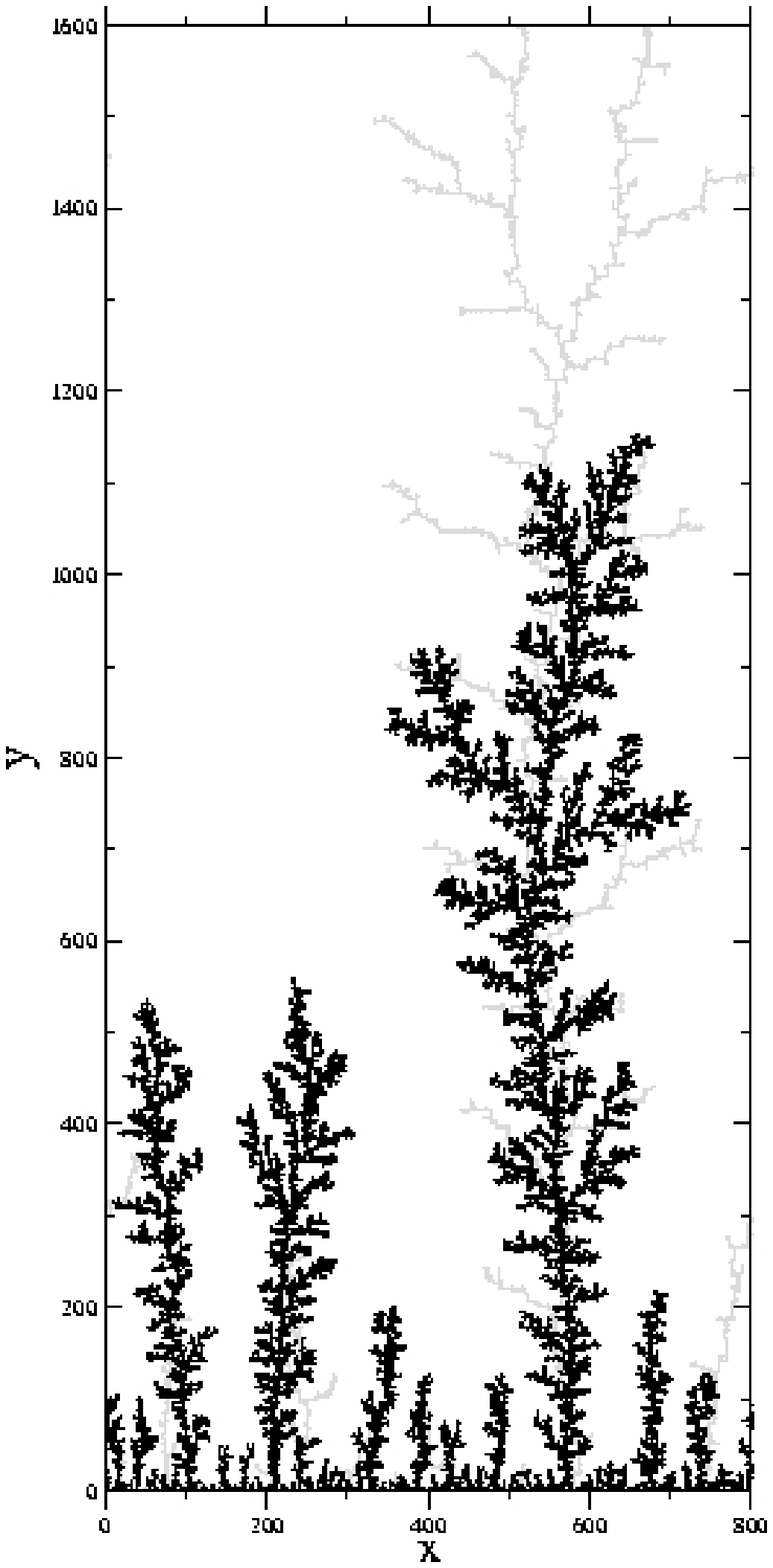}
\vspace{1cm}
\caption{Snapshots of two deposits at $T^*=10^{-1}$
(black) and $T^*=10^{-4}$ (grey).}
\label{fig1}
\end{figure}

\newpage

\begin{figure}
\includegraphics[width=7cm]{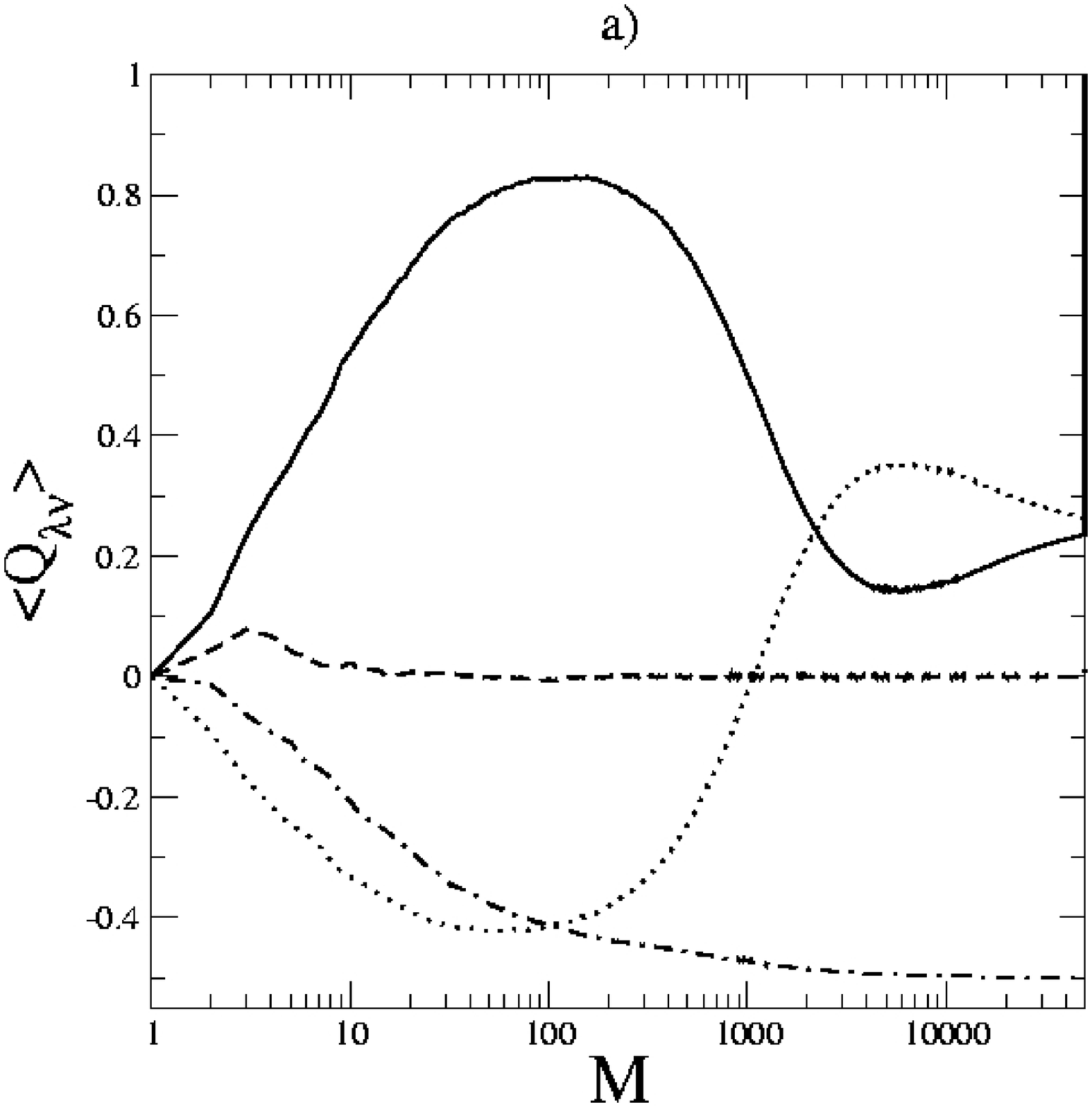}
\vspace{1.5cm}
\includegraphics[width=7cm]{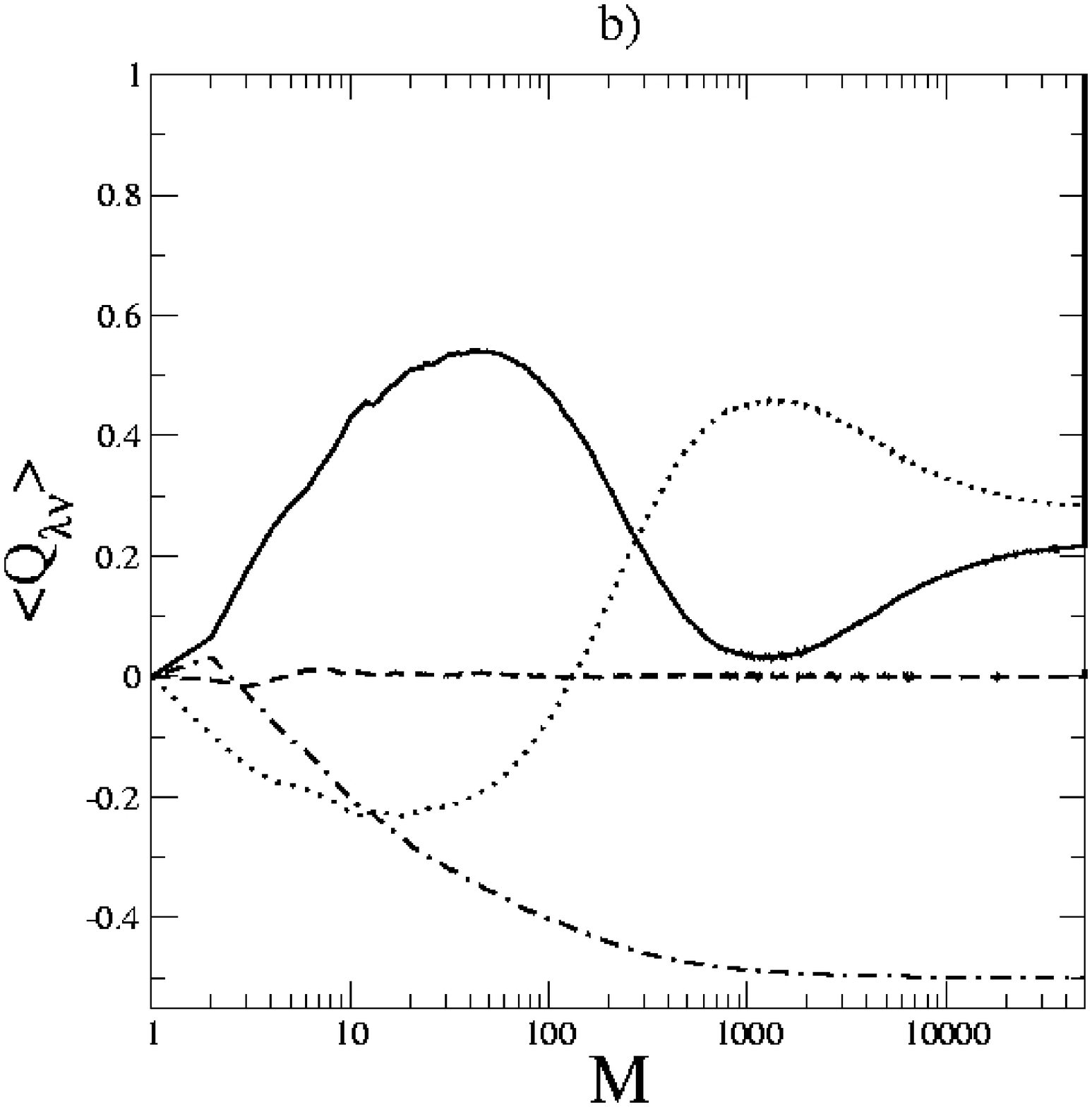}
\vspace{1cm}
\caption{Mean values of the components of the ordering matrix $\bf Q$ 
as a function of the stage of growth. We identify three
regimes: (i) initial
growth with decreasing $Q_{zz} \ne -0.5$, (ii) early growth with
decreasing $Q_{zz} \sim -0.5$ and increasing $Q_{xx}$ and decreasing
$Q_{yy}$ and (iii) late growth with $Q_{zz} = -0.5$ and increasing $Q_{xx}$
$\sim$ decreasing $Q_{yy}$ (see text for details). (a), $T^*=10^{-1}$ and (b),
$T^*=10^{-4}$. 
$Q_{xx}$ - full line; $Q_{yy}$ - dotted line; $Q_{zz}$ - dot-dashed line; 
$Q_{xy}$ - dashed line.}
\label{fig2}
\end{figure}

\newpage

\begin{figure}
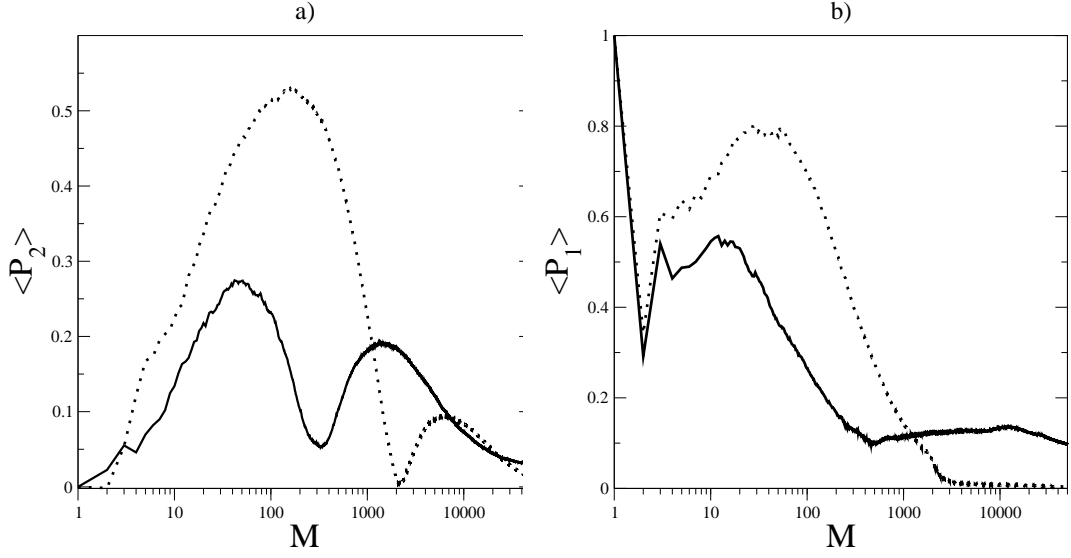

\includegraphics[width=7cm]{p2xmgr.eps}
\vspace{1.5cm}
\includegraphics[width=7cm]{p1xmgr.eps}
\vspace{1cm}
\caption{Mean values of (a) 
the nematic ($\langle P_2\rangle$) and (b) the 
ferromagnetic ($\langle P_1\rangle$) 
order parameters as a function of the 
stage of growth. Dotted line: $T*=10^{-4}$. 
Full line: $T*=10^{-4}$.}
\label{fig3}
\end{figure}

\newpage

\begin{figure}
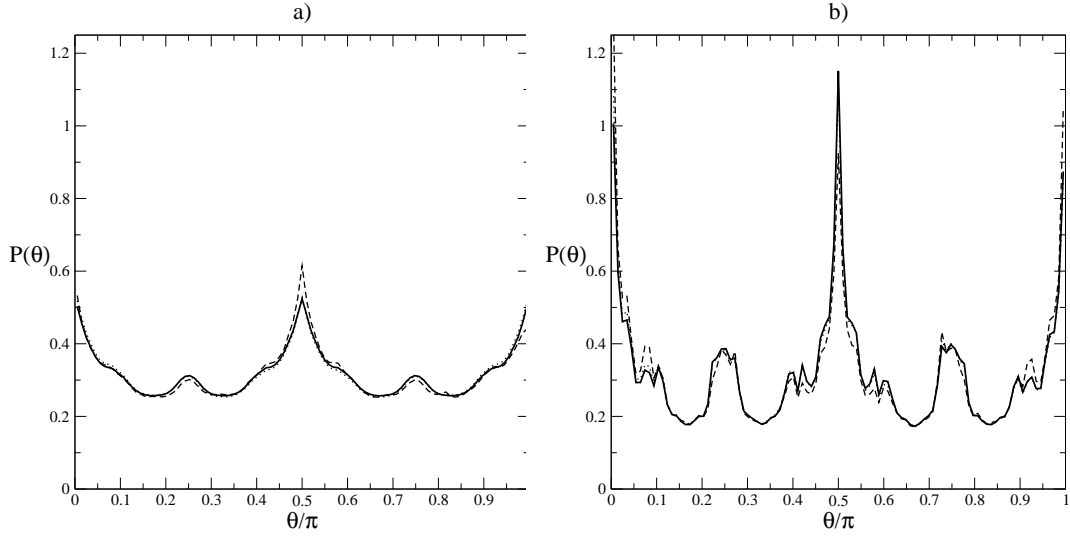

\includegraphics[width=7cm]{orcorT01xmgr.eps}
\vspace{1.5cm}
\includegraphics[width=7cm]{orcorT00001xmgr.eps}
\vspace{1cm}
\caption{Mean value of the 
orientational probability density at $T^*=10^{-1}$ (a) and at 
$T^*=10^{-4}$ (b). The different lines correspond to different 
sizes of the deposit: full line, $M=50000$; dotted line $M=10000$; 
dashed line $M=2000$.}
\label{fig4}
\end{figure}

\newpage

\begin{figure}
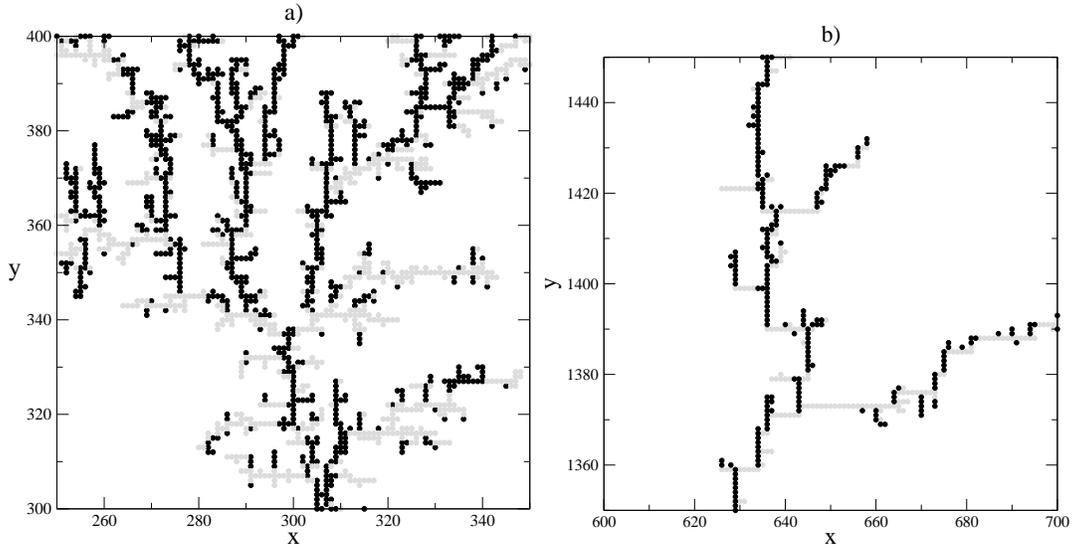

\includegraphics[width=7cm]{pol2cT01xmgr.eps}
\vspace{1.5cm}
\includegraphics[width=7cm]{polar2colorxmgr.eps}
\vspace{1cm}
\caption{Details of snapshots for two deposits at $T^*=10^{-1}$ (a) 
and at $T^*=10^{-4}$ (b). The dipoles whose absolute value of the
vertical component is larger (smaller) 
than the absolute value of the horizontal 
component are plotted in black (grey).}
\label{fig5}
\end{figure}

\newpage

\begin{figure}
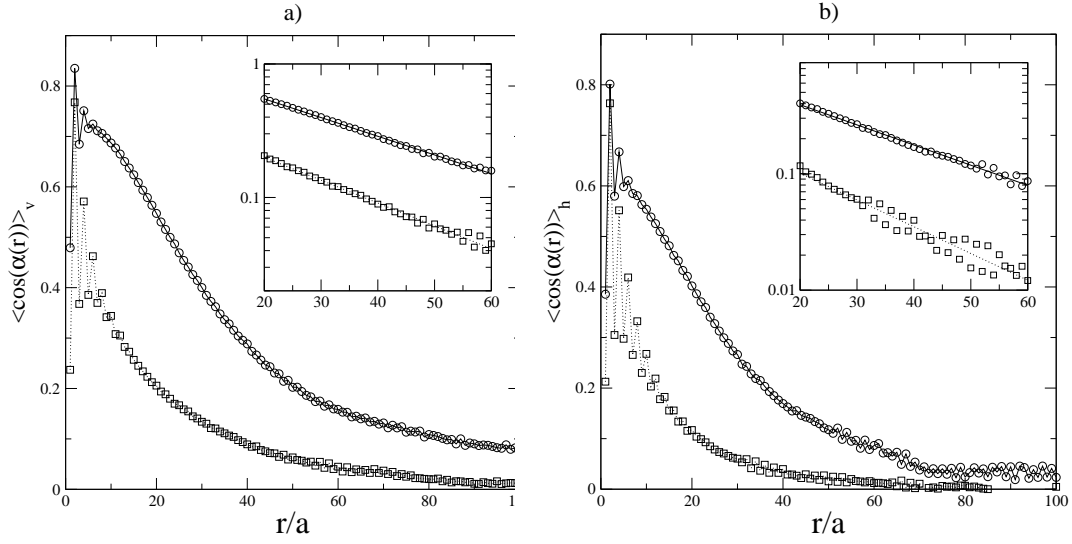

\includegraphics[width=7cm]{verangxmgr.eps}
\vspace{1.5cm}
\includegraphics[width=7cm]{horangcorrxmgr.eps}
\vspace{1cm}
\caption{Mean value of the co-sine of the angle between dipoles on the
same column (a) or row (b) a distance $r$ apart for late stage deposits,
with $M=50000$. 
The circles correspond to $T^*=10^{-4}$ and the squares to $T^*=10^{-1}$. 
The lines in the main figures are a guide to the eye. 
The lines in the insets are linear regressions illustrating the
exponential decay of correlations in that range of distances.}
\label{fig6}
\end{figure}


\begin{thebibliography}{99}

\bibitem{review} P. I. C. Teixeira, J. M. Tavares and M. M. Telo da
Gama, J. Phys.: Cond.\ Matter {\bf 12}, R411 (2000).

\bibitem{levesque:1993} J. J. Weis and D. Levesque, 
Phys.\ Rev.\ E, {\bf 48},
3728 (1993). 

\bibitem{Patey:1992} D. Wei and G. N. Patey, Phys.\ Rev.\ A, {\bf 46},
7783 (1992). 

\bibitem{levesque:1993b} J. J. Weis  and D. Levesque, 
Phys.\ Rev.\ Lett., {\bf 71},
2729 (1993).

\bibitem{Rubi:1995} R. Pastor-Satorras and J. M. Rub\'{\i},
Phys.\ Rev.\ E {\bf 51}, 5994 (1995).

\bibitem{Meakinbook} P. Meakin, {\it Fractals, scaling and growth far
from equilibrium} (Cambridge University Press, Cambridge, 1998).

\bibitem{Meakin:1983} P. Meakin, Phys.\ Rev.\ A, {\bf 27}, 2616 (1983).

\bibitem{Barabasibook} A.-L. Barab\'asi and H.E. Stanley, 
{\it Fractal concepts in surface growth}  
(Cambridge University Press, Cambridge, 1995).


\bibitem{jpcm:2002} F . de los Santos, M. Tasinkevych, J. M. Tavares and
P. I. C: Teixeira, ``Deposition of magnetic particles: 
a computer simulation study'', submitted to J. Phys.: Cond.\ Matter.

\bibitem{footnote1} This step is slightly different from that of 
\cite{Rubi:1995}. In that work only a new position is generated,  
and the orientation of the dipole is changed to the direction of the 
local field if this new position is accepted.   

\bibitem{Weis:2002} J. J. Weis, Molec.\ Phys.\ {\bf 100}, 579 (2002).


\bibitem{allenbook} M. P. Allen and D. J. Tildesley, {\it
Computer simulation of liquids}, (Oxford University Press, Oxford, 1987).

\end{thebibliography}
\end{document}